\documentclass{aa}
\usepackage{graphicx}
\usepackage{natbib}
\usepackage{amssymb}
\begin{document}

\title{Time evolution of X-ray coronal activity in PMS stars; a
possible relation with the evolution of accretion disks \thanks{Table 1 is only available in electronic form at http://www.edpsciences.org} }
\author{E. Flaccomio \and G. Micela \and  S. Sciortino}
\offprints{E. Flaccomio,
\email{ettoref@oapa.astropa.unipa.it}}

\authorrunning{Flaccomio et al.}
\titlerunning{Activity from the PMS to the ZAMS}

\institute{INAF - Osservatorio Astronomico di Palermo Giuseppe S. Vaiana,
Palazzo dei Normanni, I-90134 Palermo, Italy\\
\email{ettoref@oapa.astropa.unipa.it, giusi@oapa.astropa.unipa.it, sciorti@oapa.astropa.unipa.it}}

\date{Received 24-09-02 / accepted 06-02-03}

\abstract{We investigate the evolution of X-ray stellar activity from the age
of the youngest known star forming regions (SFR), $\lesssim 1$Myr, to about 100
Myr, i.e. the zero age main sequence (ZAMS) for a $\sim 1M_\odot$ star. We
consider five SFR of varying age ($\rho$ Ophiuchi, the Orion Nebula Cluster,
NGC~2264, Chamaeleon I, and $\eta$ Chamaeleontis) and two young clusters (the
Pleiades and NGC~2516). Optical and X-ray data for these regions are retrieved
both from archival observations and recent literature, and reanalyzed here in a
consistent manner so to minimize systematic differences in the results. 

We study trends of $L_{\rm X}$ and $L_{\rm X}/L_{\rm bol}$ as a function of
stellar mass and association age. For low mass stars ($M \lesssim 1M_\odot$) we
observe an increase in $L_{\rm X}/L_{\rm bol}$ in the first 3-4 Myr and a
subsequent leveling off at the {\em saturation} level ($L_{\rm X}/L_{\rm bol}
\sim -3$). Slowly evolving very low mass stars then retain saturated levels
down to the oldest ages here considered, while for higher mass stars activity
begins to decline at some age after $\sim 10^7$ years. 

We find our data consistent with the following tentative picture: low mass PMS
stars with no circumstellar accretion disk have saturated activity, {\em
consistently} with the activity-Rossby number relation derived for MS stars.
Accretion and/or the presence of disks somehow lowers the observed activity
levels; disk dissipation and/or the decrease of mass accretion rate in the
first few Myrs of PMS evolution is therefore responsible for the observed
increase of $L_{\rm X}/L_{\rm bol}$ with time.

} 

\maketitle

\section{Introduction \label{sect:intro}}

Since the advent of sensitive X-ray imaging observations, late type ($>$F), low
mass stars are known to be sources of X-rays with luminosities spanning more
than 4 orders of magnitude, from $L_{\rm X} =10^{27}$  or less to more than $10^{31}$
ergs/sec \citep{vai81}. Given the similarity with the solar emission (e.g.
thermal emission-line X-ray spectrum, flares) it is believed that X-ray
emission from low mass stars originates from stellar coronae that are
essentially similar to the solar one ($L_{\rm X,\odot} =0.2$-$5.0 \cdot 10^{27}$
ergs/sec, \citealt{per00}), albeit sometimes much more active. 

In order to observationally investigate the physical origin of the large
observed spread in $L_{\rm X}$, homogeneous stellar samples have been successfully
employed to restrict the number of free parameters. Open clusters and Star
Forming Regions (SFR) are ideal targets in this respect, as all stars have
approximatively the same age and metallicity. The comparison between different
regions then ideally permits to investigate the age evolution and the
metallicity dependence of X-ray activity.

For late-type slowly rotating stars of nearby field population, coronal
and chromospheric activity indicators correlate with stellar rotation
\citep{pal81} and with convective zone properties (\citealt{noy84,piz02} and
references therein).  Such correlations however do not hold for fast
rotating stars, whose observed activity levels appear to have reached {\em
saturation} (e.g. $L_{\rm X}/L_{\rm bol}\sim 10^{-3}$). X-ray activity has also been
found to decrease with increasing stellar age as stars loose angular momentum
through an activity driven magnetized wind. Such observed relations are the
main evidence that activity in the main sequence (MS) is produced by a stellar
dynamo whose efficiency depends, at least for slow rotations, on stellar
rotation and convection (the so called $\alpha-\omega$ dynamo).  

While MS activity can be convincingly  ascribed to the stellar dynamo, the
mechanism responsible for coronal activity in low mass, pre-main-sequence (PMS)
stars is not understood. During the last 20 years several close-by SFRs have
been observed in the X-rays (e.g. \citealt{fei93,dam95,fla02a,fla02b}), finding
bright X-ray emission from most PMS stars. No conclusive evidence for the
action of a MS-type dynamo mechanism has been found for these stars. Although
the presence of a convective layer does seem to be important for the existence
of an active corona (massive fully radiative PMS stars do not show coronal
emission), no relation has been generally found between the X-ray emission
level of PMS stars and their rotation periods\footnote{An exception is the
Taurus region (e.g. \citealt{stel01}). However it is also possible that the
correlation between activity and rotation is in this case the consequence of
the possibly more fundamental correlations between activity and mass-accretion
\citep{fla02b,fla02c} and between rotation and accretion(e.g.
\citealt{bou97}).} and/or convective region depth (or turn-over times).
Instead, a relation has been often found with stellar mass or bolometric
luminosity (e.g. \citealt{fei93,fei02,fla02b}), parameters unrelated to the
$\alpha-\omega$ dynamo. It is therefore not clear at present whether a new
mechanism should be invoked to explain coronal activity in the PMS (e.g. a
different dynamo), or the same dynamo acting on the MS works here in a
saturation regime, and/or additional phenomena occur in the PMS coronae that
hide the underlying correlations with the $\alpha-\omega$ dynamo parameters.

The picture is probably further complicated by matter accretion phenomena
occurring on many PMS stars through a circumstellar disk: either accretion or
the presence of the disk may play a role in determining the observed X-ray
activity levels. Several studies have indeed found that accreting stars have
lower mean X-ray activity levels, and more time-variable emission respect to
their non-accreting counterparts \citep{neu95,fla00,fla02b,fla02c,stel01}.
However, this point is controversial: other authors have failed to detect a
significant  difference in activity levels of accreting and non accreting stars
(e.g. \citealt{fei02,pre02,fei93}). As discussed by \citet{fla02c}, we believe
that the discrepancy in the results is due, rather than to differences in the
data themselves, to the different sensitivity of the techniques used
to compare the activity levels of the two classes. In this work we will
therefore assume that accretion disks do influence activity.  The physical
nature of such influence is presently not understood. Some hypotheses, proposed
by \citet{dam95} and \citet{fla02b}, speculate that accretion (and/or the
associated outflow) may modify the geometry of the magnetic field that confines
the coronal plasma, either reducing the fraction of the stellar surface
available for closed magnetic loops, or modifying more radically the magnetic
field, maybe with the involvement of the magnetized inner parts of the
circumstellar disk (e.g. \citealt{mon00}).

While the study of individual SFRs and open clusters allows us to study the
correlation (or lack of correlation) of X-ray activity with parameters such as
stellar rotation, mass and circumstellar accretion indicators, important clues
to the mechanism responsible for activity are expected to come from the
comparison of SFR at different evolutionary stages. 

In this work we assume, to first approximation, that the differences in
activity between stars belonging to different associations can be attributed
exclusively to their different evolutionary stages. In other words, we assume
that, in different associations, stars of a given mass are statistically
indistinguishable at the time of their formation (at least in regard to
activity) and that their evolution is not influenced by different ambient
conditions. These assumptions may be questioned because, for example:
differences in initial angular momentum and stellar metallicity may affect
X-ray activity; differences in environment, such as stellar density, can affect
properties of the stellar population, such as the frequency of multiple systems
and the circumstellar disk lifetime \citep{hai01b} which may both influence
observed X-ray activity levels. Here, neglecting these possible complications,
we will directly compare results obtained for stellar associations of varying
age, in an attempt to investigate the effect of stellar evolution on activity.
More precisely, we will select stars in definite mass ranges and we will follow
the evolution with time of two activity indicators, $L_{\rm X}$ and $L_{\rm X}/L_{\rm bol}$. 

In order to complete our program, we consider here six young PMS stellar groups
with ages ranging from $\sim$ 1 to $\sim$ 7 Myrs: $\rho$ Ophiuchi, the Orion
Nebula Cluster (ONC), split in the central and outer regions (see below),
NGC~2264, the Chamaeleon I association and $\eta$ Chamaeleontis. In addition we
consider two older ZAMS clusters: the Pleiades and NGC~2516. These regions have
been selected so to probe a significant age range and because optical and X-ray
data suitable for our analysis were available either in the literature or in
data archives. In all cases we try to consider stellar samples that are likely
free from X-ray selection effects and, when needed, we re-analyze the published
data in an homogeneous way. The latter is a crucial step because different
instrumentation, assumptions and analysis techniques employed to derive
physical quantities from both X-ray and optical observations can result in
significant systematic differences that would interfere with our purpose. 

In the following we first introduce the procedures employed to consistently
analyze (or re-analyze) the data for the various regions (Sect. 
\ref{sect:conc_ana}); we then discuss each of them individually, the sources of
our data, the analysis performed, and the results obtained in regard to the
dependence of activity on stellar mass (Sect.  \ref{sect:conc_SFR}). Finally we
present our evolutionary picture (Sect.  \ref{sect:conc_XvsA}) and propose a simple
physical model that tries to explain our observational findings (Sect. 
\ref{sect:conc_disc}). 

\section{Generals of data analysis \label{sect:conc_ana}}

The quantities we will deal with are two X-ray activity indicators ($L_{\rm X}$ and
$L_{\rm X}/L_{\rm bol}$), stellar masses and ages. The estimates of these quantities are
often subject to significant uncertainties, due both to errors in the
observational quantities and to the assumptions made in order to convert these
latter into physical parameters. In order to make the comparisons of activity
among different associations meaningful, special care must be therefore taken
on the one hand to reduce random uncertainties and on the other to minimize
systematic differences due to the non-uniform approaches and assumptions made
in the different studies. In the following two subsections we review the steps
taken to this effect.

\subsection{X-ray luminosities \label{sect:genLx}} 

We utilize data obtained with four X-ray detectors: {\em ROSAT} HRI and PSPC,
{\em Chandra} HRC-I and ACIS-I. In the original works from which these data are
taken, X-ray luminosities were calculated from measured instrumental
count-rates assuming different spectral bands, the choice depending largely on
the spectral response of the instrument. These X-ray luminosities are not
directly comparable. The count-rate to luminosity conversion factors were
computed in all cases assuming optically thin isothermal plasma emission. The
assumed plasma temperatures ($kT$), however,  varied from case to case, as well
as the way in which interstellar extinction was accounted for. Here we
recompute all conversion factors, using PIMMS\footnote{Portable, Interactive,
Multi-Mission Simulator}, for a single spectral band (0.1-4.0 keV). For PMS
stars we assume $kT = 2.16$ keV, consistent with recent results on several SFRs
(e.g. \citealt{fei02,fla02a,ima01a}), and an hydrogen column density, $N_{\rm H}$,
proportional to optical/IR extinction, $A_{\rm V}$ or $A_{\rm J}$, where the latter is
measured, with the exception of $\eta$ Chamaeleontis (Sect.  \ref{sect:eta_cha}),
individually for each  member of the association.

For the two MS clusters, NGC~2516 and the Pleiades, we assume (cf.
\citealt{gag95b}) lower temperatures, between $\sim 0.4$ and $\sim 1.0$keV.
Given the small and quite uniform interstellar extinction, $N_{\rm H}$ was assumed,
for stars lacking individual measurements, to correspond to the mean $A_{\rm V}$ of
the entire cluster. 

Departure from our assumptions for $kT$ and $N_{\rm H}$ will result in errors in the
count rate to $L_{\rm X}$ conversion, whose magnitudes depend on the detector used
for the observation and on the range of variation of the two parameters for the
given stellar population. In the following we will discuss, for some of the
regions considered, the sensitivity of the conversion factor to these
assumptions. In all cases the likely systematic uncertainty in $L_{\rm X}$ is smaller
than $\sim 0.1$ dex.

\subsection{Masses and Ages \label{sect:conc_gen_mass}} 

We make use of \citet[][ SDF hereafter]{sie00} evolutionary tracks\footnote{See
also the web site: 
http://www-laog.obs.ujf-grenoble.fr/activites/starevol/evol.html.} in order to
estimates masses and ages of PMS stars from effective temperatures and
bolometric luminosities. These latter values are taken from published works,
derived in most cases from medium resolution optical or IR spectra. For the 
two oldest PMS regions, Chamaeleon I and $\eta$ Chamaeleontis, estimates are
based on optical/IR photometry. Ages for the two ZAMS clusters are taken from
\citet{mey93} and are  derived from the turnover point in the upper main
sequence; we then estimate stellar masses from optical photometry and the
theoretical isochrones appropriate to the clusters age (SDF; solar
metallicity), transformed from the theoretical $L_{\rm bol}$ - $T_{\rm eff}$ plane into
the observational  $V$ - $(B-V)$ plane with the relations given by
\citet{ken95}.  Note that masses for both PMS and MS stars are estimated from
the same theoretical evolutionary model. In order to alleviate the effect of
residual uncertainties in the mass estimates (e.g. due to the different
temperature scales used in the original studies the data are taken from) we
will consider wide mass ranges, similarly to what done for the ONC by
\citet{fla02b}. 

\begin{figure}[t]
\centering
\resizebox{\hsize}{!}{\includegraphics{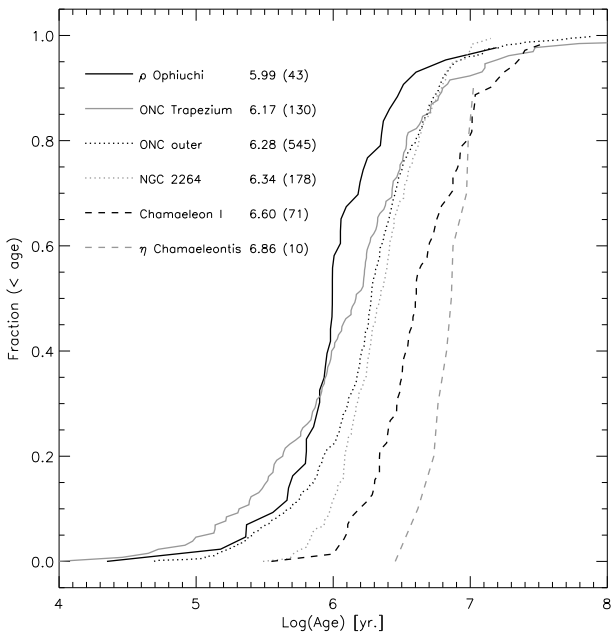}}
\caption{Cumulative age distribution for the six PMS regions under study. The legend on the upper-left side reports, for each region, the median
of the logarithm of the age (in years) and, in parenthesis the number of
stars in the sample.}
\label{fig:conc_SFR_age}
\end{figure}

We now turn to the problem of PMS stellar ages: individual ages inferred from
PMS evolutionary models are known to be subject to large uncertainties. In
addition to the uncertainties in the theoretical evolutionary models,
bolometric luminosities are subject to an artificial spread due to a number of
effects \citep{har01}. As a result the real dispersion of stellar ages in a
given star forming region may be much smaller than estimated from its HR
diagram. We decided to characterize each PMS stellar group with a single age,
i.e. the median of the ages derived from the SDF tracks. As long as the shape
of the age distribution due to the artificial spread in $L_{\rm bol}$ is at least
similar for all star forming regions, we can assume that the median of the {\em
measured} ages is in some close and unique relation with the {\em real} median
age. Figure \ref{fig:conc_SFR_age} shows the distribution of individual stellar
ages, as derived in the following section, for each of the six PMS stellar
groups considered here. Also indicated are the median values which will be used
in the following.

\section{The star forming regions and open clusters \label{sect:conc_SFR}}

We now discuss the eight stellar groups taken in consideration, describing the
stellar samples and the analyses performed, along the lines sketched in the
previous section, in order to make the results comparable to each other. We
proceed in chronological order, starting from the youngest stellar association,
according to the sequence suggested by Fig.  \ref{fig:conc_SFR_age}.

\subsection{$\rho$ Ophiuchi}

$\rho$ Ophiuchi, with an age of $\lesssim 10^6$ years, is the youngest region
here considered. It is a medium size association forming out of a dense
molecular cloud, and most of its members are deeply embedded in the cloud. It
has been therefore studied with more success in the near IR and X-ray bands
(e.g. \citealt{luh99,ima01a}). Our source of optical/IR data for $\rho$
Ophiuchi members is the work by \citet{luh99} who obtained NIR spectroscopy and
photometry of 114 stars and derived, for 69 of these, effective temperatures, J
band extinctions and bolometric luminosities. We restrict our analysis to the
field of view (FOV) of the two {\em Chandra} X-ray observations described
below, so lowering the census of our well characterized members (i.e. with
$T_{\rm eff}$ and $L_{\rm bol}$) to 51. We estimated masses and ages for 43 of these
stars using SDF evolutionary tracks, the remaining eight having apparently
masses lower than the lowest mass for which SDF tracks are available
($0.1M_\odot$). These 43 stars comprise our reference sample for $\rho$
Ophiuchi.

\subsubsection{X-ray data and analysis \label{sect:conc_oph_x}}
 
We have retrieved, from the {\em Chandra} Data Archive ($\rm
http://asc.harvard.edu/cda/$), two deep ($\approx 100$ ksec each) and slightly
overlapping ACIS-I observation of the $\rho$ Ophiuchi cloud (Obs. Id. 635,
637). A thorough analysis of the first observation has been already reported by
\citet{ima01a}, focusing mainly on medium resolution X-ray spectra and on
source variability; \citet{ima01b} have used both observations to study X-ray
emission of the known brown dwarfs and candidates brown dwarfs in the region.
However, \citet{ima01b} do not produce a  list of sources and \citet{ima01a} do
not compute  upper limits to X-ray luminosities of undetected cluster members,
a necessary step in order to avoid selection effects. To uniform our analysis
to the general principles stated in Sect.  \ref{sect:intro}, we re-analyzed the two
dataset.

First, in order to selectively lower the background level we filtered the two
event files retaining only events with energies in the 0.3 to 8.0 keV range,
likely containing the largest fraction of the X-ray flux from $\rho$ Ophiuchi
members. We then detected sources and measured count rates\footnote{Effective
exposure times were taken from an exposure map computed using the tools in the
{\em Chandra Interactive Analysis of Observations} ({\sc ciao}) package ($\rm
http://asc.harvard.edu/ciao/$), for an incident energy of 2.16 keV. The choice
of energy is however not critical as vignetting depends only mildly on energy
for the range of energies relevant to stellar coronae.} using {\sc pwdetect}
(\citealt{dam97}, Damiani et al., in preparation)\footnote{See also
http://www.astropa.unipa.it/progetti\_ricerca/PWDetect} and a signal to noise
ratio threshold of 5.10, which, given the background level of the two
observations, is expected to yield $\sim 1$ spurious detection in each FOV.
Ninety sources were detected in obs. 635 and 77 in obs. 637, after removal from
the latter source list of $\sim$20 spurious detection due to a bright CCD
column in coincidence with a luminous X-ray source. For comparison
\citet{ima01a} detect 87 sources in obs. 635. Seven sources located in the
small overlap region of the two observations ($\sim 40$arcmin$^2$), were
detected in both observations. In the same region 6 sources appeared instead in
only one of the two datasets. We created a merged source list, assigning to the
sources detected in both observations uncertainty-weighted averages of count
rates and spatial coordinates. Our merged source list comprises 160 distinct
X-ray sources with count rates ranging from $6.3\cdot 10^{-5}$ to 0.79 $\rm
cnts \cdot s^{-1}$.  

We cross-identified our detected X-ray sources with the list of 43 selected
$\rho$ Ophiuchi members discussed in the previous section. For comparison
purposes (see below), we have also identified, with the same list of members,
the X-ray sources detected by \citet{ima01a}. In both cases the identification
process went smoothly, thanks to the high spatial resolution of {\em Chandra}
and the relatively low spatial density of objects in the region. We identify 36
members with 35 X-ray sources, a close pair of members (GY240A and GY240B)
being not resolved in the X-ray image (we treat the measured count-rate of the
pair as an upper limit to the count-rate of each star). For the seven members
undetected in the ACIS-I data we computed upper limits to count rates using
{\sc pwdetect} and the same signal to noise ratio threshold used for
detections. 

We estimate X-ray luminosities of stars in our sample from ACIS-I count rates
and assuming a distance of 165pc \citep{dame87}.\footnote{There is some
uncertainty on this distance: an {\em Hipparcos} based estimate of the distance
to the Upper Scorpius region, of which the $\rho$ Ophiuchi cloud is part,
indicates $\sim$145pc \citep{dez99}. Had we adopted this value $L_{\rm X}$ would be
lowered by $\sim 0.12$dex. The $L_{\rm X}/L_{\rm bol}$ values, on which our main
conclusions are based would remain unaffected. Note also that the median
stellar age of the region members would increase by $\sim$0.17dex, becoming
almost equal to that of the ONC Trapezium region (Sect.  \ref{sect:ONC}).}
Count rates to flux conversion factors were computed, as described in Sect. 
\ref{sect:genLx}, i.e. taking $kT=2.16$keV and $N_{\rm H}$ proportional to the $J$
band absorption, $A_{\rm J}$ (see below). Thanks to the analysis of medium resolution
ACIS-I spectra performed by \cite{ima01a}, which also yielded independent
estimates of X-ray luminosities, we are in the position to check these
assumptions.

The median(mean) of the quiescent state $kT$ derived from spectral fits
\citep{ima01a} for 28 stars in our member list is 2.1(2.3) keV, indeed similar
to our value. Moreover, considering the relevant uncertainties of these best
fit $kT$, their distribution is consistent with a narrow range of values:
$\sim$50\% fall inside the  $kT=1.5-2.5$ keV range and another 35\% have 90\%
error bars consistent with the same interval. 

\begin{figure}[t]
\centering
\resizebox{\hsize}{!}{\includegraphics{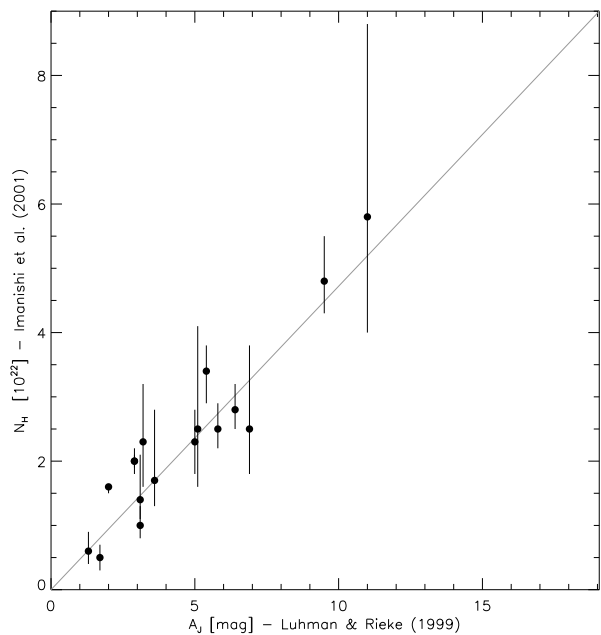}}
\caption{$N_{\rm H}$ derived from fittings of X-ray spectra \citep{ima01a} vs. $J$ band extinction, $A_{\rm J}$, measured for the same stars by \citet{luh99}. The straight line corresponds to the median of the $N_{\rm H}/A_{\rm J}$ ratios: $4.72\cdot 10^{21}$.  Error bars are 90\% confidence intervals derived from spectral fits.
\label{fig:conc_oph_nh_cmp}}
\end{figure}

In regard to extinction, Fig.  \ref{fig:conc_oph_nh_cmp} shows the
proportionality between X-ray derived $N_{\rm H}$ and $A_{\rm J}$. The proportionality
constant was assumed to be the median of the measured $N_{\rm H}/A_{\rm J}$: $4.72 \cdot
10^{21}$. This is somewhat lower than what would be expected on the basis of a
normal optical/IR extinction curve \citep{mat90}, $A_{\rm V}/A_{\rm J}$=3.55, and the
relation between  $N_{\rm H}$ and $A_{\rm V}$ given either by \citet{ryt96}, $N_{\rm H}=2.0\cdot
10^{21}A_{\rm V}$ or \citet{pred95}, $N_{\rm H}=1.79\cdot 10^{21}A_{\rm V}$. With these
assumptions we would have derived $N_{\rm H}/A_{\rm J}=(N_{\rm H}/A_{\rm V}) (A_{\rm V}/A_{\rm J}) = 7.1 \cdot
10^{21}$ and $6.35\cdot 10^{21}$ respectively, i.e. about 30-50\% higher than
our value.  This would have in turn resulted in overestimating the  conversion
factors (and $L_{\rm X}$) by 0.1-0.2 dex, depending on absorption in the $A_{\rm V}=2$-$50$
range. The departure from the {\em normal} relation may points toward a
peculiarity of the extinction law in the $\rho$ Ophiuchi cloud (cf.
\citealt{vrb93,ken98}). 

\begin{figure}[t]
\centering
\resizebox{\hsize}{!}{\includegraphics{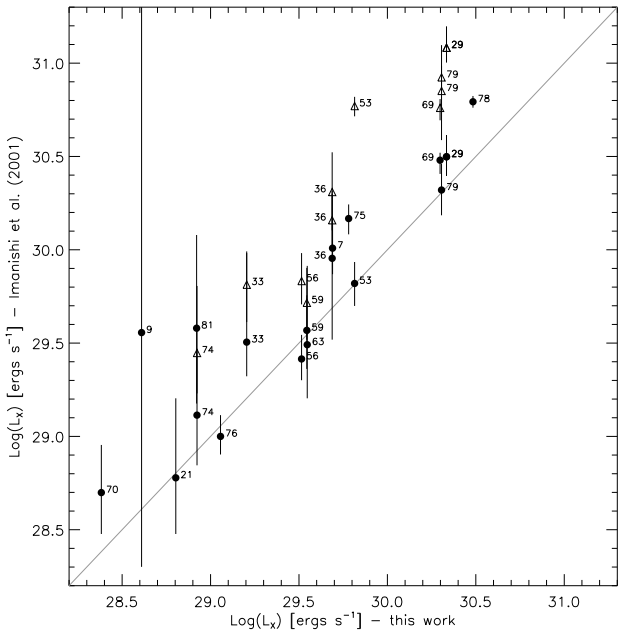}}
\caption{Comparison of X-ray luminosities estimated from the same ACIS-I data in this work and by \citet{ima01a}. Filled circles
indicate that the $L_{\rm X}$ estimate by \citet{ima01a} refer to the quiescent
source emission, while empty triangles refer to flares. The numbers beside each point
is the source identification number given by \citet{ima01a}. Note that up to
three estimates (one quiescent and two flares) are given for each source. Error
bars are 90\% confidence intervals derived from spectral fits.}
\label{fig:conc_oph_lx_cmp}
\end{figure}

Finally, Fig.  \ref{fig:conc_oph_lx_cmp} compares our X-ray luminosities with
those estimated by \citet{ima01a} on the basis of spectral fittings; 90\%
confidence interval are also shown for the latter. More than one estimate of
$L_{\rm X}$ is given by \citet{ima01a} for some variable sources corresponding to
quiescent and flaring states (indicated with different symbols in the Figure)
while we only derive an averaged $L_{\rm X}$ over the whole observations. In general,
the estimates for quiescent $L_{\rm X}$ by \citet{ima01a} and our estimates are fully
compatible, with the possible exception of some {\em flaring} sources, for
which the difference can be easily attributed to the presence of the flare. 

We conclude that, at least for $\rho$ Ophiuchi stars, the assumption of a
single plasma temperature for all sources and of optical extinctions to
calculate conversion factors yields realistic values of $L_{\rm X}$. From the
dispersion of spectral-fit $kT$ and $N_{\rm H}$ about our assumed values, we estimate
that the mean error on our conversion factors is of the order of $0.1$ dex.

\subsubsection{Activity vs. mass \label{sect:conc_oph_res}}

We study the relation between activity and stellar mass using the same method
described in \citet{fla02b} for the Orion Nebula Cluster. Figure
\ref{fig:conc_lx_mass_pms} shows X-ray luminosity functions for the stars in
five mass bins. The same Figure shows the median $L_{\rm X}$ and the dispersion (25
and 75\% quantiles) of the luminosity functions as a function of mass. Figure
\ref{fig:conc_lxlb_mass_pms} shows the same kind of analysis for our second
activity indicator: $L_{\rm X}/L_{\rm bol}$. We draw the following conclusions: 1) the
median $L_{\rm X}$ is correlated with mass; 2) the median $L_{\rm X}/L_{\rm bol}$
($10^{-4.5}$-$10^{-4.0}$) is significantly below the canonical saturation
level, $10^{-3}$; 3) $L_{\rm X}/L_{\rm bol}$ is correlated with mass, at least at for $M
\lesssim 1M_\odot$. Note that the correlation of $L_{\rm X}/L_{\rm bol}$ with mass was
also found in the ONC by \citet{fla02b}: here however the dependence appears to
be somewhat steeper.

\subsection{Orion Nebula Cluster \label{sect:ONC}}

The X-ray ({\em Chandra} HRC) data for the ONC are described in detail in
\citet{fla02a}. There we also defined an optically selected, extinction limited
sample of well characterized ONC members (there referred to as the {\em optical
sample}) that we also use here.

For the present purpose, i.e. the study of the evolution of activity in the
PMS, we split the region surveyed by the HRC observation in two sub-areas: the
central $5' \times 5'$, roughly identifiable with the so called {\em Trapezium}
region, and the remaining of our HRC FOV. It has already been noted (e.g.
\citealt{hil97}) that the Trapezium region is of somewhat more recent formation
respect to the outer part of the molecular cloud. Figure \ref{fig:conc_SFR_age}
shows the cumulative age distribution for the two regions: it is evident that,
although the two distributions overlap there is an excess of very young stars
in the Trapezium. Part of the overlap may be due, other than to the artificial
and real age spread in the two regions, to the fact that the area we identify
with the Trapezium, likely contains, for a projection effect, a number of stars
that are in fact located in front of the Trapezium and should belong more
properly to the outer region. In any case, following the discussion in Sect. 
\ref{sect:conc_ana} we will assume that the {\em median} activity levels of
stars in these two regions are representative of the two {\em median} stellar
ages.

Figure \ref{fig:conc_lx_mass_pms} and \ref{fig:conc_lxlb_mass_pms} show our
{\em standard}  analysis of activity vs. mass for the two spatially segregated
regions. We note that the same qualitative trends found for the whole ONC by
\citet{fla02b} are retrieved in each of the two sub-regions.

\subsection{NGC~2264}

A recent analysis of stellar activity in NGC~2264 was presented by
\citet{fla02c}; these authors utilized published X-ray and optical data
\citep{fla00,reb02b} to investigate the relation between circumstellar
environment and activity. The data were there reanalyzed following the general
criteria stated in Sect.  \ref{sect:conc_ana}. Here we use the same data and the
same reference stellar sample. The sample comprises 178 candidate members from
the  spectroscopic sample of \citet{reb02b}, to which \citet{fla02c} were able
to assign masses, ages and X-ray luminosities (or upper limits), these latter
derived from the {\em ROSAT} HRI count rates listed by \citet{fla00}. 

Figure \ref{fig:conc_lx_mass_pms} and \ref{fig:conc_lxlb_mass_pms} show our
analysis of $L_{\rm X}$ and $L_{\rm X}/L_{\rm bol}$ vs. stellar mass. Given the large number of
upper limits, it is quite clear that the sensitivity of the presently available
X-ray data ({\em ROSAT} HRI from \citealt{fla00}) is not sufficient to define
the low luminosity tails of the X-ray luminosity functions so that the median
$L_{\rm X}$ is defined only in two mass bins and the trend of $L_{\rm X}$ with mass seen
for the other regions is not so apparent here. The median $L_{\rm X}/L_{\rm bol}$ shows a
pattern similar to the other regions, being close to the saturation level for
$M\lesssim 1.0M_{\rm \odot}$ and declining for larger masses.

\subsection{Chamaeleon I}

The Chamaeleon I association, with an age of 5-6 Myrs, is the second oldest
star forming region considered in this study.  Like for NGC~2264, the optical
and X-ray data we use here are described in detail by \citet{fla02c}, where the
activity/disk connection was analyzed. The original data that work is based on
were taken for the most part from \citet{law96}, but reanalyzed consistently
with what done for the other SFRs.

Figure \ref{fig:conc_lx_mass_pms} shows the X-ray luminosity functions in our
chosen mass ranges and the  $L_{\rm X}$ vs. mass scatter plot. Figure 
\ref{fig:conc_lxlb_mass_pms} refers, in the same format, to $L_{\rm X}/L_{\rm bol}$. The
general trend of increasing $L_{\rm X}$ with increasing mass, seen in the other star
forming regions,  can be clearly observed here. $L_{\rm X}/L_{\rm bol}$ seems to be close
to the saturation level at all masses. In particular we note that, respect to
the younger association, $\rho$ Ophiuchi and the ONC, we do not observe  a
decrease of the median $L_{\rm X}/L_{\rm bol}$ at very low masses.

\subsection{$\eta$ Chamaeleontis \label{sect:eta_cha}}

$\eta$ Chamaeleontis is a small, nearby (97 pc) and relatively old ($\sim
10$Myr) PMS cluster discovered through {\em ROSAT} X-ray observations by
\citet{mam99}. The X-ray properties of 12 members have been investigated by
\citet{mam00}, while \citet{law01} report optical photometry and derive
rotational periods. The total $\eta$ Chamaeleontis known population counts 15
stars: 3 are early type stars (one of which, HD75505 is undetected in the
X-rays), 11 are WTTS and 1 is a CTTS. Ten of the 12 low mass members have been
discovered through detection in the X-rays \citep{mam99}. The following optical
search for additional members conducted by \citet{law02} resulted in the
discovery of only two non X-ray selected members, among which the CTTS (ECHA
J0843.3-7905). 

In the following, reassured by the high detection efficiency among the known
low-mass  cluster population, we will only consider X-ray detected members. The
X-ray ({\em ROSAT} HRI) count rates are taken from \citet{mam00} while the
optical data, $T_{\rm eff}$ and $L_{\rm bol}$, come from \citet{law01}. Following the
previous investigations we assumed zero absorption along the line of sight. The
determination of masses, ages and X-ray luminosities followed the principles
stated in Sect.  \ref{sect:conc_ana}.

\begin{figure*}[!t!]
\resizebox{17cm}{!}{\includegraphics{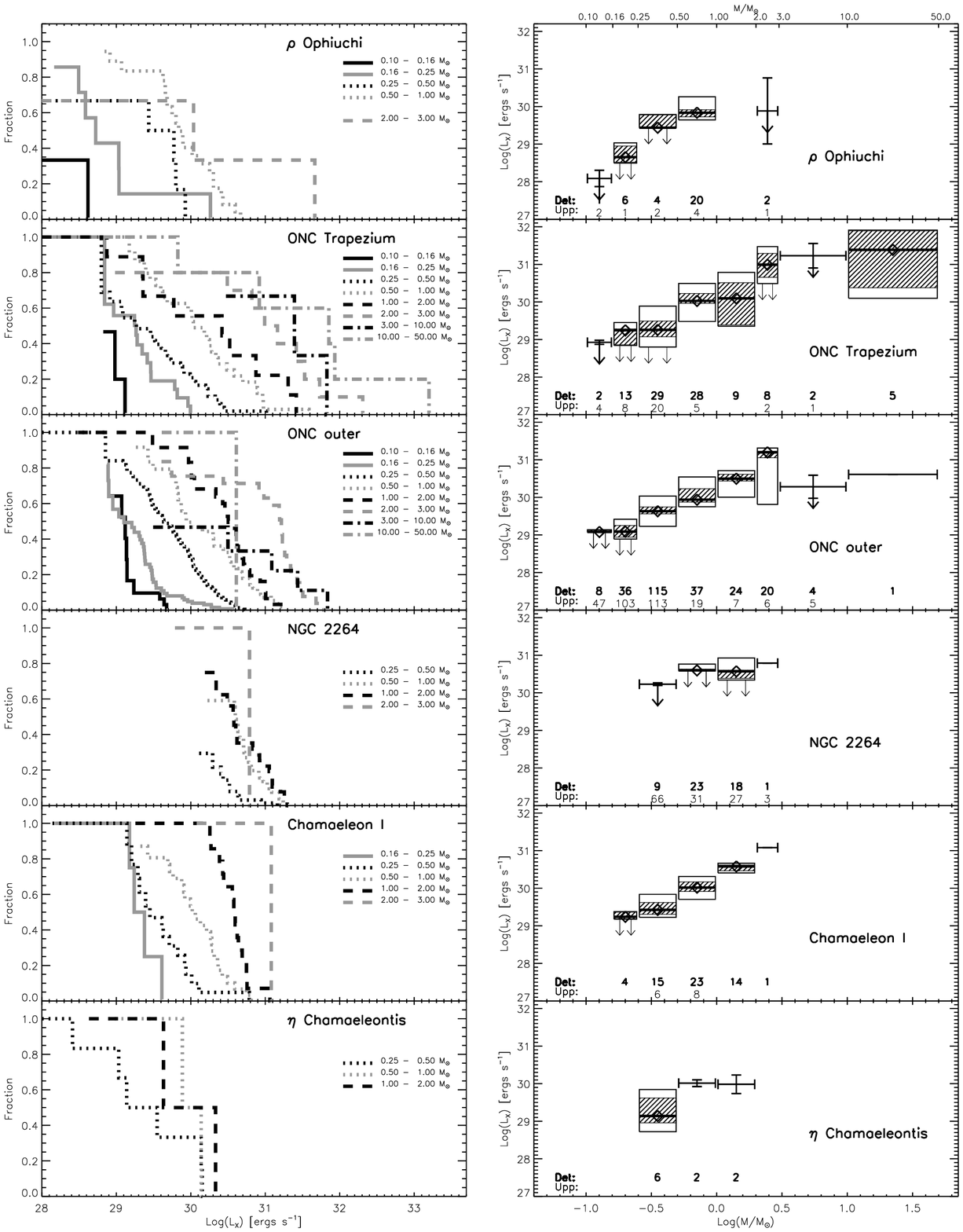}}
\caption{{\sc Left}: X-ray luminosity functions (XLF) for our PMS associations
in several mass ranges. {\sc Right}: Median and scatter of  $\log(L_{\rm X})$ as a function of
$\log(M/M_{\rm \odot})$. Medians are indicated by diamonds
centered on thick segments, with uncertainties indicated by the shaded boxes, and the 25\% and
75\% quantiles of the XLF by thin-lined boxes. 
Two small downward-pointing arrows at box bottom indicate an upper limit on the
lower error in median and/or the lower quartile. In some cases medians are substituted by averages (either measured or upper limits), indicated by error bars.
Numeric values above abscissa
for each box indicate numbers of detections (``Det.'') and upper limits
(``Upp.'')  contributing to the XLFs shown on the left. \label{fig:conc_lx_mass_pms}}
\end{figure*}

\begin{figure*}[!h!]
\resizebox*{17cm}{!}{\includegraphics{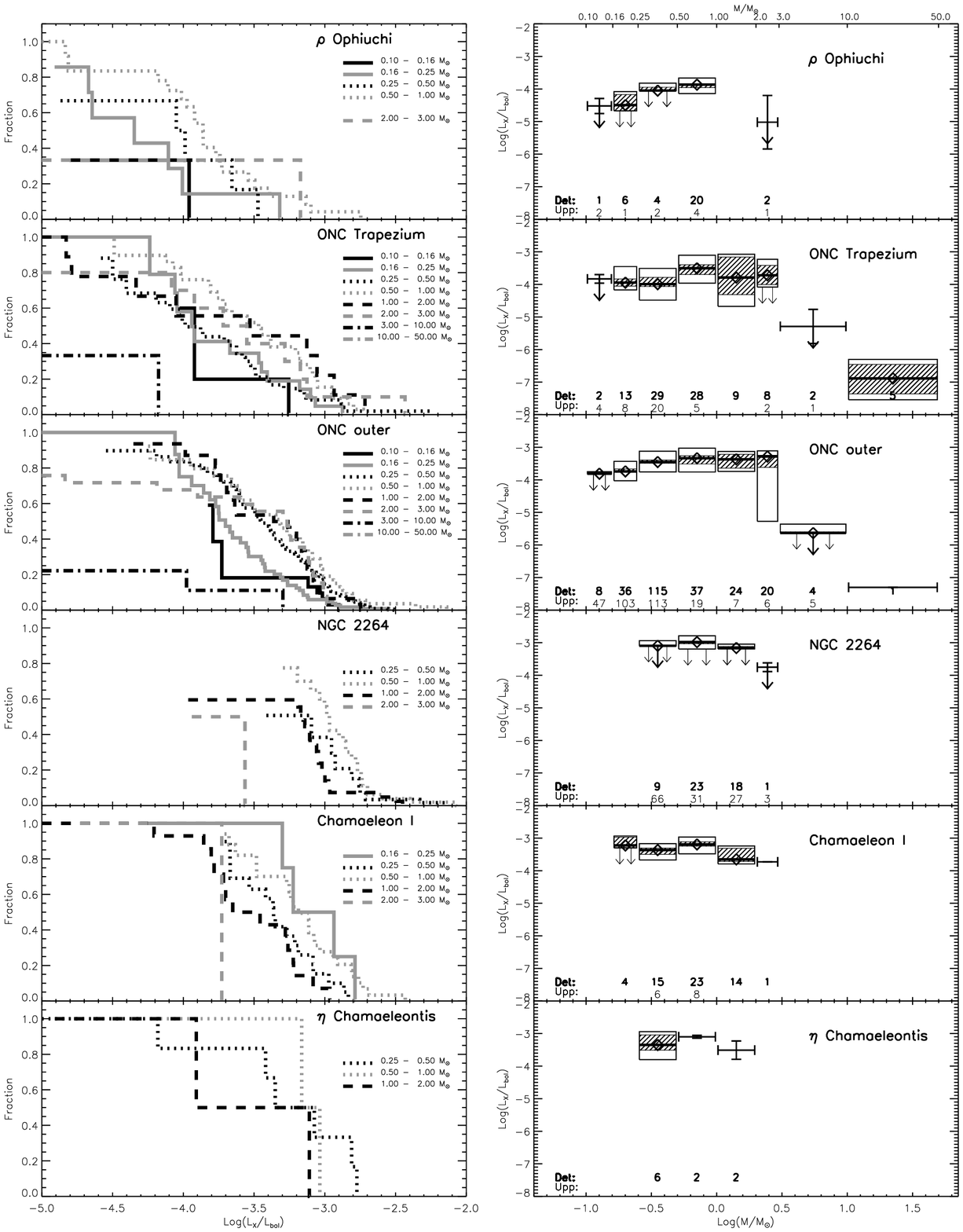}}
\caption{Same as Figure \ref{fig:conc_lx_mass_pms} but for $L_{\rm X}/L_{\rm bol}$ instead of $L_{\rm X}$. In the scatter plots, diamonds with downward-pointing arrows indicate upper limits to the medians. \label{fig:conc_lxlb_mass_pms}} 
\end{figure*}

Figure \ref{fig:conc_lx_mass_pms} shows the X-ray luminosity functions in our
mass ranges and the trend of median $L_{\rm X}$ with mass. Figure 
\ref{fig:conc_lxlb_mass_pms} refers, in the same format, to $L_{\rm X}/L_{\rm bol}$. We
note that, notwithstanding the low number of stars, the data points are
compatible with the same qualitative results found for low mass stars in the
slightly younger Chamaeleon I region: a dependence of $L_{\rm X}$ on mass and a
constant 'saturated' $L_{\rm X}/L_{\rm bol}$.

\subsection{The Pleiades \label{sect:conc_pleiades}}

The Pleiades open cluster, with an age of  $\sim 100$Myr \citep{mey93} and a
distance of only $\approx 127$pc, is possibly the best studied young ZAMS
cluster in the solar neighborhood. Its stellar population has been nearly
completely identified and several X-ray studies performed with {\em ROSAT}
(e.g. \citealt{sta94,mic96,mic99}) have measured X-ray luminosities for a
representative fraction of Pleiades members. Here we use this wealth of
information for the purpose of extending our study of the evolution of activity
from the pre-main-sequence to the ZAMS.


Our sample of Pleiades members includes $\sim 500$ stars listed in the open
cluster database of C.F. Prosser and J.R. Stauffer\footnote{ 
http://cfa-www.harvard.edu/$\sim$stauffer/opencl/index.html} and that fall in
the FOV of the X-ray observations described below. We estimate stellar masses
from the theoretical SDF isochrone and $V_0$ magnitudes as explained in  Sect. 
\ref{sect:conc_gen_mass}. The choice of using the $V$ magnitude instead of, for
example, the $B-V$ color as a mass indicator is due to the fact that $V$ is
available for all the stars in our sample (vs. $\sim 54\%$ for $B-V$). We
exclude from the following analysis 38 stars that, in the $V_0$ vs. $(B-V)_0$
diagram, are more distant than 1 magnitude from the isochrone (along the $V_0$
axis). We note that, given the non-perfect adherence of the observational
member locus to the SDF isochrone and to the presence of unresolved multiple
systems (having lower $V_0$ than those pertaining to the primary alone) our
mass estimates are somewhat uncertain. We get a rough idea of the uncertainties
on mass estimates by comparing our masses derived from $V_0$ with those
obtained from the same isochrone through $(B-V)_0$. The difference in the
individual masses are typically smaller than 0.05 dex, small compared to the
width of the mass ranges used to compute the Maximum Likelihood distributions
of $L_{\rm X}$ and $L_{\rm X}/L_{\rm bol}$.

Bolometric luminosities, used for computing $L_{\rm X}/L_{\rm bol}$, are estimated, in
the same way as masses, from the theoretical isochrone.


X-ray data are taken from the published analysis of 16 different X-ray
exposures performed with {\em ROSAT}, 8 using the PSPC \citet{mic98} and 8
using the HRI \citep{mic99}. Each star in our reference catalog was observed
between one and nine times ($\sim 3$ times on average). However, possibly due
to sensitivity limits, only 46\% of the stars in our reference list are
detected, while for the remaining ones only upper limits to the X-ray
luminosities are available.

The estimates of X-ray luminosities we use are taken from \citet{mic98} for the
PSPC data and \citet{mic99} for the HRI data. Both refer to the 0.1-2.5keV
spectral band and are based on the following assumptions: a coronal temperature
$kT=0.8$keV and absorption $N_{\rm H}=2\cdot10^{21} \cdot 3.1 \cdot E(B-V)$, where
$E(B-V)$ is measured individually for about 15\% of the stars, while for the
remaining ones a mean value, 0.04, is used.  The spectral band luminosities
refer to is narrower than our {\em standard} 0.1-4.0keV band; anyway, because
of the low coronal temperature of Pleiades sources, the difference in
conversion factor due to the difference in band is small: for $kT=0.8$keV, the
fraction of emitted energy falling in the 2.5-4.0keV band, and therefore the
amount by which we underestimate $L_{\rm X}$, is $\lesssim 0.01$ dex. Only for  $kT
\sim 1.5$ keV, hotter than typical MS coronal sources \citep{gag95b,dan02}, the
difference in $L_{\rm X}$ due to the narrower spectral band reaches $\sim 0.05$ dex. 
Softer spectra, $kT \sim 0.4-0.5$keV, have been inferred for main sequence $F$
type stars \citep{gag95b,fla02,dan02}: assuming $kT=0.8$keV results in this
case in an overestimation of $L_{\rm X}$ of up to $\sim 0.05$dex, which is small
compared to the other sources of random uncertainties (e.g. counting statistic
and distance spread). The assumption of an average value of  absorption, for
stars for which no individual measurement of $E(B-V)$ is available, is also
justified: 80\% of the stars for which $E(B-V)$ has been measured  have a
corresponding $N_{\rm H} \lesssim 10^{21}$ $\rm cm^{-2}$, which, for  $kT=0.8$ keV
translates into a difference on the individual conversion factor of $\lesssim
0.1$dex respect to assuming the mean absorption. Moreover, while stars that
have individual extinction estimates are located in a sky area of high
nebulosity (this being the reason that prompted the measurement of individual
extinctions), the remaining stars are located in an area apparently devoid of
clouds and most likely have fairly uniform $E(B-V)$; the former estimate of 0.1
dex for the uncertainty on the conversion factor is therefore most likely too
pessimistic.

Given the multiplicity of observations, in order to assign a single $L_{\rm X}$ value
to each of our stars, we proceeded as in \citet{fla00}, i.e. we compute maximum
likelihood mean luminosities using all the available information (detections
and upper limits). This was implemented by computing, with the ASURV package
\citep{fei85}, the Kaplan Mayer estimator of the values of $\log(L_{\rm X})$ available
for each source. Because of the time variable nature of X-ray emission, the thus
derived mean $\log(L_{\rm X})$ of $\sim 6\%$ of the stars which were detected in some
of the observations turned out to be upper limits. Out of the 483 stars in
our reference sample we therefore have estimates of mean $\log(L_{\rm X})$ for 40\%
of the stars and upper limits for the remaining ones.  

Figure \ref{fig:conc_lx_mass_ms} shows X-ray luminosity functions in our chosen
mass bins and the $\log(L_{\rm X})$ vs. mass scatter plot. The same analysis is
repeated for $\log(L_{\rm X}/L_{\rm bol})$ in Fig.  \ref{fig:conc_lxlb_mass_ms}. Low mass
stars ($\log(M/M_{\rm \odot}) \lesssim -0.3$) appear to be saturated, with
$\log(L_{\rm X}/L_{\rm bol}) \sim -3$, and show the consequent mean correlation of $L_{\rm X}$
with mass. Higher mass stars however are not saturated, so that the relation of
luminosity with mass is lost, as also evinced by the onset of the activity -
rotation connection which is indeed observed at this mass boundary
\citep[cf.][]{mic99}.

\subsection{NGC~2516 \label{sect:conc_2516}}

The open cluster NGC~2516 is in many respects similar to the Pleiades, and it
has indeed been often named the southern Pleiades. It is however slightly older,
$\sim 140$ Myrs  \citep{mey93}, and more distant (387 pc, \citealt{jef97}).

Our data are based on the photometric survey of \citet{jef01} and the analysis
of eight {\em Chandra} X-ray observations reported by  \citet{dam02}. Our
initial member list is the same as that used by \citet{dam02} and comprises
$725$ stars indicated as members by \citet{jef01} and falling in the FOV of at
least one of the {\em Chandra} observations. Stellar masses, and bolometric
luminosities, were derived analogously with what described for the Pleiades
cluster from $V_0$ magnitudes and a theoretical (SDF) isochrone, the only
difference being in the choice of the isochrone appropriate to the larger age
of NGC~2516. We excluded from our reference sample 52 stars that, in the $V_0$
vs. $(B-V)_0$ diagram, lie farther away than 1 mag (along the $V_0$ axis) from
the 140 Myr isochrone and four stars that were found to be outside the $M_V$
range covered by the isochrone.

We adopt X-ray luminosities from \citet{dam02}. These are averages of the
luminosities estimated from detected source counts in all the analyzed
observations. Conversion factors between counts and luminosity in the
0.1-4.0keV  band are there computed assuming uniform extinction, $N_{\rm H}=8.2\cdot
10^{20}$, and, for bright sources, a two temperature spectral model (one of
which fixed at 0.26keV), constrained by observed hardness ratios. For weaker
stellar sources and upper limits, the conversion factor is taken as the median
of the values derived for bright sources with the same spectral type.
\citet{dam02} estimate that, given the range of variation of the conversion
factors derived from hardness-ratios, the uncertainty in {\em individual}
$\log(L_{\rm X})$, introduced by an inaccurate spectral model, is at most $\sim 0.15$
dex.

\begin{figure*}[!t!]
\resizebox{17cm}{!}{\includegraphics{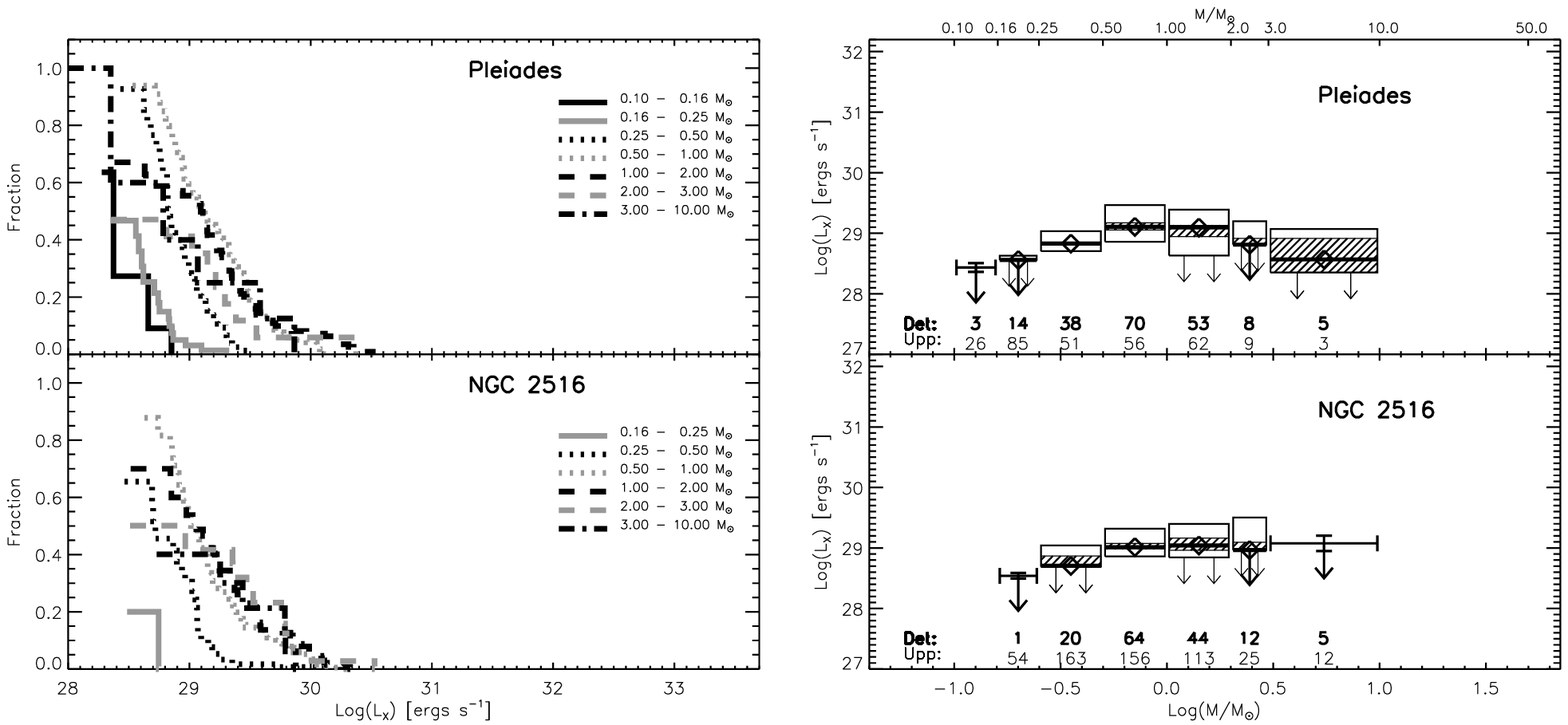}}
\caption{Same as Figure  \ref{fig:conc_lx_mass_pms} for the two ZAMS clusters.  \label{fig:conc_lx_mass_ms}} 
\end{figure*}

\begin{figure*}[!t!]
\resizebox{17cm}{!}{\includegraphics{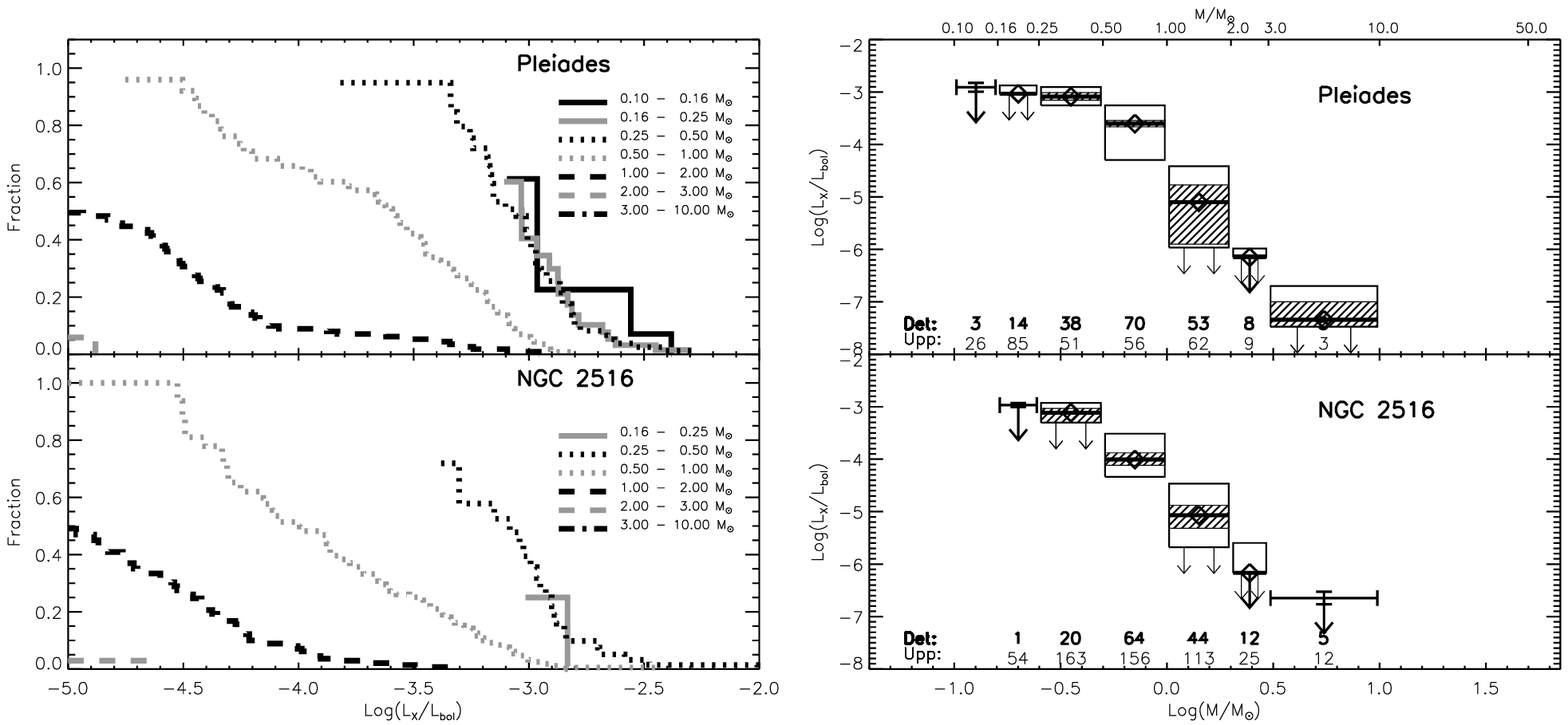}}
\caption{Same as Figure \ref{fig:conc_lxlb_mass_pms} for the two ZAMS clusters. \label{fig:conc_lxlb_mass_ms}} 
\end{figure*}

Figure \ref{fig:conc_lx_mass_ms} and \ref{fig:conc_lxlb_mass_ms} show our
standard analysis of activity, $L_{\rm X}$ and $L_{\rm X}/L_{\rm bol}$, as a function of mass. The same qualitative trends observed for the Pleiades are retrieved here.


\section{Evolution of coronal activity \label{sect:conc_XvsA}}

We now utilize the data presented in the previous section to study the
evolution of stellar activity in specific stellar mass ranges. 
Table \ref{tab:data}, available in electronic format, summarizes the data described in Sect.  \ref{sect:conc_SFR} and also utilized in the following.
For our purpose (see Sect.  \ref{sect:conc_gen_mass}) we assume that the {\em
median} activity level of each roughly coeval PMS stellar group is
representative of the PMS activity at the {\em median} of stellar ages in the
same group. 

\begin{table}
\caption{Optical and X-ray data for the stellar sample}
\begin{tabular}{cccccc}
\hline

N.& Region& Mass& $\rm \log(Age)$& $\rm \log(L_{\rm X})$& $\rm \log(L_{\rm X}/L_{\rm bol})$ \\
& & [$\rm M_\odot$]& [yr.]& $\rm [ergs \ s^{-1}]$&  \\
\hline
(1)&(2)&(3)&(4)&(5)&(6)\\
\hline
\end{tabular}
\label{tab:data}
\begin{list}{}{}
\item[Notes -- ] Column (1): sequential reference number; column (2): identifier of the region the star belongs to. The table, comprising 2129 rows, is available in electronic form at: http://www.astropa.unipa.it/$\sim$ettoref/Act\_Evol\_T1.dat
\end{list}
\end{table}

\begin{figure*}[!t!]
\resizebox{17cm}{!}{\includegraphics{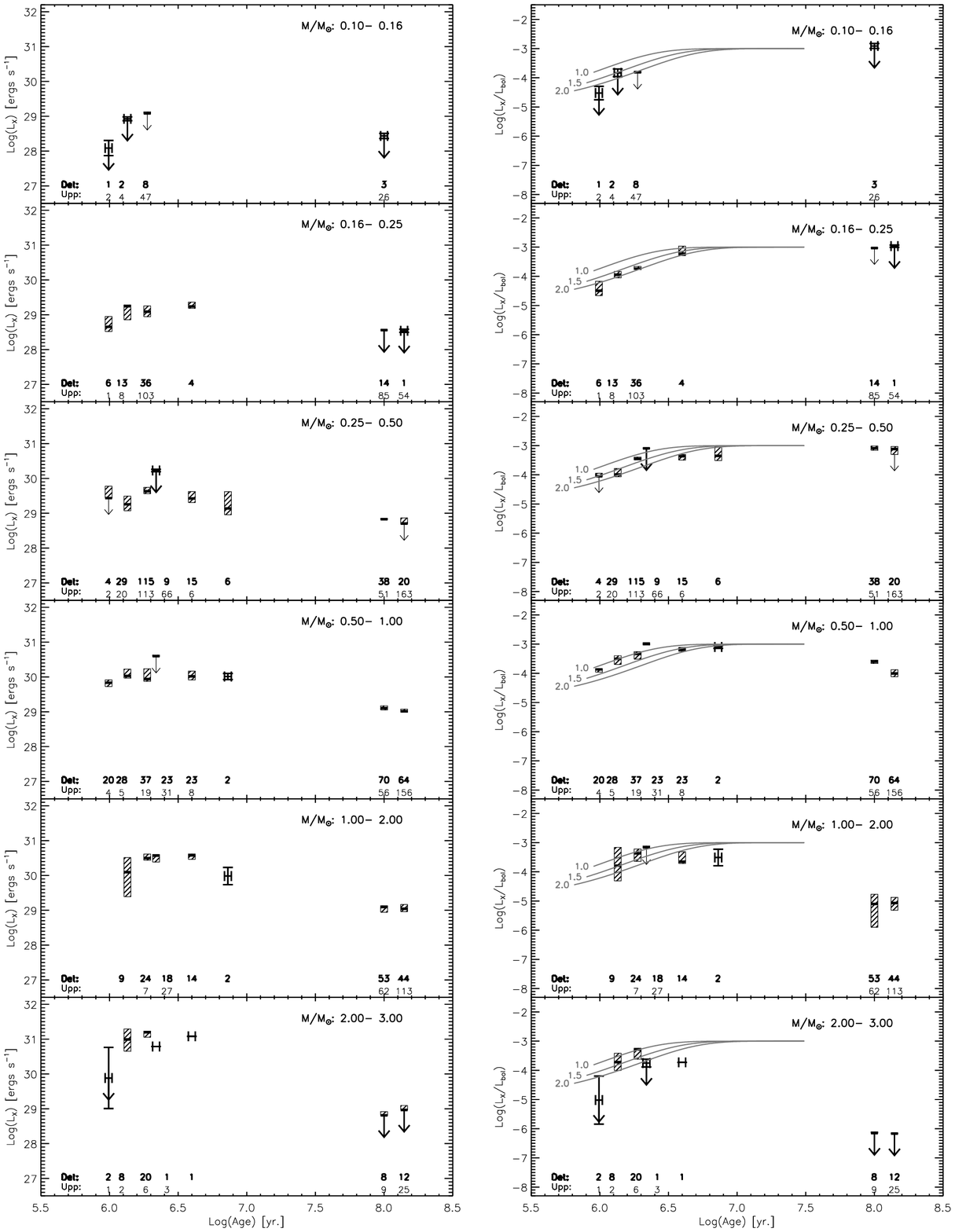}}
\caption{Time evolution of median $L_{\rm X}$ and $L_{\rm X}/L_{\rm bol}$ (left and right respectively). Medians are represented by small horizontal segments, with hatched boxes representing uncertainties. Down-pointing arrows indicate upper limits on the median and on the lower error (thick and solid arrows respectively). In some cases medians are substituted with averages (either defined or upper limits), indicated by error bars. Numeric values above abscissa indicate numbers of detections (``Det.'') and upper limits
(``Upp.'') in each of the bin. The three curve in the right hand panels refer to three realizations of the simple model described in Sect.  \ref{sect:conc_disc}, characterized by the disk dissipation time, $\tau_{\rm disk}$, whose value is reported beside the curves in units of $10^6$ years.  \label{fig:lx_lxlb_age}} 
\end{figure*}

Figure \ref{fig:lx_lxlb_age} shows, in the left hand column, for each of our
chosen mass ranges, the median $\log(L_{\rm X})$ of each of the eight regions studied
in the previous section as a function of median association age. The same
analysis is presented for $\log(L_{\rm X}/L_{\rm bol})$ in the right hand column. In order
to assess the significance of the difference between $L_{\rm X}$ and $L_{\rm X}/L_{\rm bol}$
distributions of stars in a given mass range and belonging to different regions
(i.e. having different ages) we have performed two population statistical tests
using the ASURV package. ASURV performs five different tests that give
quantitatively different results. In comparing distribution of activity
indicators we will consider, quite arbitrarily, {\em significantly} and {\em
possibly} different two samples for which the majority of the tests give lower
than, respectively, 1\% and 10\% probability that the samples are not drawn
from the same parent population. In order to further assess the presence of
correlations between median activity and age of the PMS regions we have also
performed two correlation tests (namely the Cox proportional hazard model test
and the generalized Kendall`s $\tau$ test), also using ASURV.

Summarizing the trends for $\log(L_{\rm X})$ on the left side of Fig. 
\ref{fig:lx_lxlb_age} as well as the results of the tests, we have weak
evidence of an increase of $\log(L_{\rm X})$ in the PMS, for $M<0.25M_\odot$. More
specifically, in the $M=0.16-0.25 M_\odot$ mass bin, $\rho$ Ophiuchi
($\log(Age)=5.99$) {\em may} have lower levels respect to the outer ONC
($\log(Age)=6.28$) and the Chamaeleon I region ($\log(Age)=6.60$). In the
following two mass ranges there is no evidence for a temporal evolution of
median $L_{\rm X}$, with the notable exception of NGC~2264 ($\log(Age)=6.34$),
for which the medians, or the high luminosity tail of the luminosity functions
if the median are not defined, are higher than that of the other regions of
similar age (significantly so in the 0.5-1.0 $M_\odot$ range). Similar lack of
evidence for a PMS evolution is retrieved for the two highest mass bins,
although the two population tests indicate a significant lowering of $L_{\rm X}$
between the age of Chamaeleon I and $\eta$ Chamaeleontis (the two oldest PMS
regions) in the 1-2 $M_\odot$ range. In all mass bins the results of the
correlation tests do not evidence any overall trends of $L_{\rm X}$ with age in the
PMS. A significant decrease of $L_{\rm X}$ from the PMS to the age of the two ZAMS
clusters is observed at all masses. 

The trends of $\log(L_{\rm X}/L_{\rm bol})$ are somewhat more definite. Apart from the
lowest mass bin, for which the evidence from the two population tests is only
at the 90\% level, we observe, for $M < 1M_\odot$, a significant trend of
increasing $\log(L_{\rm X}/L_{\rm bol})$ in the early PMS phase followed by a subsequent
leveling at the saturation level, $\sim -3$. This is clearly indicated both by
the two population tests and by the correlation tests (with the exception of
the $0.5-1.0M_\odot$ for which the result of the correlation tests are only at
the $\sim 85\%$ level, increasing to 97\% if we only consider the four youngest
regions). The saturation level is likely retained down to $\sim 100$My for $M <
0.5-1.0M_\odot$, while at higher masses we observe a decline respect to our
oldest PMS region. In the two most massive bins all the tests do not evidence a
statistically significant evolution of PMS activity as measured by
$L_{\rm X}/L_{\rm bol}$.

\section{Discussion \label{sect:conc_disc}}

We now attempt a physical interpretation of our observational results. More
specifically we discuss whether the observed activity levels can be described
with the same ideas that are successful for more evolved MS stars, i.e. with
the mechanism of a stellar dynamo driven by stellar rotation plus convection. 

A recent observational study of the dependence of activity on rotation and
convection in the MS, has been performed by \citet{piz02}, who propose two
possible {\em descriptions} of stellar activity levels. 

The first scenario relates the observed X-ray luminosity ($L_{\rm X}$, in
ergs/sec) with the stellar rotational period ($P_{\rm rot}$, in days):

\begin{equation}
L_{\rm X} \approx 10^{30} \ P_{\rm rot}^{-2}
\label{eq:concP}
\end{equation}

This relation is valid, within uncertainties, for rotational periods such that
$L_{\rm X}/L_{\rm bol} \lesssim 10^{-3}$ (where $L_{\rm bol}$ depends, on the MS, on stellar
mass), while for shorter $P_{\rm rot}$ the relation saturates at $L_{\rm X} \approx
10^{-3} L_{\rm bol}$. Stars of all masses therefore follow the same relation of
$L_{\rm X}$ with $P_{\rm rot}$ when activity is not saturated but saturate at different
values of $L_{\rm X}$ according to their bolometric luminosities. In this description
convection does not enter explicitly, although the {\em presence} of convection
is necessary for the presence of coronal activity, i.e. the relation holds only
for stars that possess a convective envelope. 

The second description involves the Rossby number, i.e. the dimensionless ratio
between the stellar rotational period and the convective turnover time:
$R_{\rm 0}=P_{\rm rot}/\tau_{\rm conv}$. The observational data, for stars in a fairly wide
range of masses, can be adequately described by the relation:

\begin{equation}
\frac{L_{\rm X}}{L_{\rm bol}} \propto R_{\rm 0}^{-2}
\label{eq:concR0}
\end{equation}

This holds for large enough values of $R_{\rm 0}$ and saturates, at $L_{\rm X}/L_{\rm bol} \sim
10^{-3}$ for smaller Rossby numbers, i.e. for short rotational periods or long
convective turnover times. In this description, like in the previous, stars of
all mass follow the same relation in the non-saturated regime; however,
contrary to the previous case, here both the saturation level and the value of
the independent variable, $R_{\rm 0}$, at which saturation occurs are independent
from mass. We therefore reach an unifying picture at the expense of the
introduction of two additional variables, $L_{\rm bol}$ and $\tau_{\rm conv}$.

As shown by \citet{piz02} the two descriptions are, in the MS phase, equivalent
(within observational uncertainties): the theoretical stellar structure models
from which the values of $\tau_{\rm conv}$ can be derived indicate that, for MS
stars in a wide range of mass, the relation  $\tau_{\rm conv} \propto
L_{\rm bol}^{-1/2}$ holds, at least approximately, with a proportionality constant
that is quite independent of stellar mass. Equation \ref{eq:concR0} is then
approximately equivalent to Eq.  \ref{eq:concP}. By considering only MS
stars, it is therefore difficult to establish which of the two relations gives
a more accurate description of the data and thus, to shed light on the physical
mechanisms responsible for activity {\em and} for the saturation phenomena.

In the following we compare these two descriptions to our PMS data. We will try
to decide whether any of the two can explain the observations, either entirely
or by introducing some ad hoc modification.

Following \citet{piz02}, we will refer to the PMS evolutionary models of
\citet{ven98}; these models compute the convective turnover time, whose
evolution in the PMS is essential for our study. Treatment of convection is
notoriously one of the most difficult points of stellar structure models:
\citet{ven98} use two alternative convective theories/approximations: Mixing
Length Theory (MLT) and Full Spectrum of Turbulence (FST). For each set of
models they compute two differently defined values of $\tau_{\rm conv}$: one
computed at the bottom of the convective zone (not defined for fully convective
stars), the other throughout the whole convective zone.  The absolute values of
$\tau_{\rm conv}$ as a function of stellar mass and age depend on its definition
and on the choice of convective theory (MLT or FST). However, the trends with
mass and age are similar in all cases. In the following, for our argumentation,
we will be mainly interested in the qualitative trends of  $\tau_{\rm conv}$
relative to the MS values, so that our conclusions do not depend on the
particular model and/or definition. First of all we note that the inverse
relation between $L_{\rm bol}$  (directly related to mass at a given age) and
$\tau_{\rm conv}$ which holds in the MS does {\em not} extend unmodified to the
PMS. Indeed, according to the \citet{ven98} models, moving to younger ages, a
star of a given mass increases both its $L_{\rm bol}$ and its $\tau_{\rm conv}$, and
this latter is more than one order of magnitude larger, for $1M_{\rm \odot}$ stars,
at the age of the PMS regions considered in this paper, respect to the MS
value. The two equations,  \ref{eq:concP} and \ref{eq:concR0}, that in the MS
are equivalent, will then produce in this case different and identifiable
predictions.

\subsection{Implications for PMS stars}

We now discuss the implications of the two descriptions of activity summarized
above for our PMS clusters, with particular reference to the ONC case, for
which a rich sample of rotational periods is available (cf. \citealt{her03}).
We recall that no relation between activity and rotation has been observed to
date in the ONC \citep{fei02,fla02b} as well as in the other regions discussed
in this work.

The most direct consequence of the our first scenario (cf. Eq. 
\ref{eq:concP}) when applied to PMS stars is that, due to the larger bolometric
luminosity of these latter respect to MS stars, the saturation level is
predicted to be reached at higher values of $L_{\rm X}$ and smaller rotational
periods. Stars with measured rotational periods  in the ONC sample ($P_{\rm rot}$
up to $\sim 20$ days) should therefore {\em not} be saturated.  This however
contrast with the absence of a $L_{\rm X}$ - $P_{\rm rot}$ correlation and with the high,
close to saturation, values of $L_{\rm X}/L_{\rm bol}$ observed in all our star forming
regions and, in particular, for stars more massive than $0.5M_{\rm \odot}$.  For
example, excluding the young $\rho$ Ophiuchi members, more than half of the
stars in the $0.5-1.0M_{\rm \odot}$ mass range of each of our star forming regions
have $\log(L_{\rm X}/L_{\rm bol}) > -3.5$. Using Eq.  \ref{eq:concP} and the $L_{\rm bol}$
values predicted by the SDF evolutionary tracks, we calculate that
$\log(L_{\rm X}/L_{\rm bol}) > -3.5$ would imply, for a $M=1.0M_{\rm \odot}$ star at the age
of the ONC, rotational periods shorter than 0.8 days. Likewise, saturated stars
($\log(L_{\rm X}/L_{\rm bol}) = -3.0$) should have $P_{\rm rot} < 0.4$. However, $\sim$97\%
and  $\sim$99\% of ONC stars have measured periods longer than 0.8 and 0.4 days
respectively. We can then exclude that in the PMS $L_{\rm X}$ is related to $P_{\rm rot}$
by Eq.  \ref{eq:concP}.

We now turn to Eq.  \ref{eq:concR0} and to the possible role of the
convective turnover time, $\tau_{\rm conv}$. As already noted the \citet{ven98}
models predict that at the age of the ONC and younger, the $\tau_{\rm conv}$ of
solar mass stars is more than 1 order of magnitude larger respect to the MS
value. Assuming that Eq.  \ref{eq:concR0} holds and that saturation occurs
at the same $R_{\rm 0}$ as for MS stars, and recalling that MS solar mass stars
saturate at $\log(P_{\rm rot})\sim 0.2$ (with $P_{\rm rot}$ measured in days, cf.
\citealt{piz02}), an increase of an order of magnitude in $\tau_{\rm conv}$
translate in an increase of the same amount in the rotational periods at which 
saturation occurs. We therefore expect PMS solar mass stars with $\log(P_{\rm rot})
\lesssim 1.2$ to be saturated. Given the similarity of convective turnover
times of stars of different masses at young ages, saturation periods of the
same order are expected for lower mass stars. Now, $\log(P_{\rm rot}) = 1.2$
corresponds to periods of about 16 days, i.e. {\em longer} than typically
measured in the ONC. This picture would therefore predict that virtually {\em
all} PMS stars are saturated. Therefore they should not not exhibit a relation
of activity vs. rotation  and they should all have $\log(L_{\rm X}/L_{\rm bol})\sim -3$.
Such a prediction is {\em almost} consistent with the observations: we do not
indeed observe a correlation of activity with rotation and for most star
forming regions we do observe almost saturated activity levels. This is
especially true for the older PMS regions (NGC~2264, the Chamaeleon I and
$\eta$ Chamaeleontis associations) and for the more massive stars (cf. Fig. 
\ref{fig:conc_lxlb_mass_pms} and \ref{fig:lx_lxlb_age}). However, quite
clearly, the youngest stars, especially at low masses, show somewhat lower
$L_{\rm X}/L_{\rm bol}$ levels. For the ONC sample of rotational periods we have no
evidence that rotational periods of low mass stars are longer than for more
massive stars. On the contrary there is likely an opposite dependence of median
$P_{\rm rot}$ with mass, the median $\log(P_{\rm rot})$ being $\sim 0.4$ (from 15 stars)
and $\sim 0.9$ (from 49 stars) in the 0.1-0.16$M_{\rm \odot}$  and
0.5-1.0$M_{\rm \odot}$ mass range, respectively.  We therefore conclude that,
although Eq.  \ref{eq:concR0}, i.e. the relation between $L_{\rm X}/L_{\rm bol}$ and
$R_{\rm 0}$, does better than the relation between $L_{\rm X}$ and $P_{\rm rot}$ (Eq.  
\ref{eq:concP}), it does not {\em by itself} explain the PMS observations.

One tentative explanation for the low $L_{\rm X}/L_{\rm bol}$ stars may invoke the
still controversial existence of the super-saturation regime \citep{ran00,
fei02}. There is some evidence of a third regime in the $L_{\rm X}/L_{\rm bol}$ vs.
Rossby number relation obtained using MS stars, other than the {\em normal}
inverse relation at large $R_{\rm 0}$ and the saturation regime: activity of stars
with very low Rossby numbers, i.e. very fast rotations or long convective
turn-over times may show a direct correlation with $R_{\rm 0}$. The faster average
rotation of lower mass  stars in the ONC, which indeed have lower average
$L_{\rm X}/L_{\rm bol}$, seems consistent with such an explanation. However, low mass
stars in the ONC span the whole range of measured rotational periods and
activity is not correlated, either directly or inversely, with rotational
period (cf. Fig.  2 in \citealt{fla02b}); this holds even when the analysis is
restricted to narrow mass ranges in order to reduce any possible difference in
stellar structure (i.e. $\tau_{\rm conv}$). Moreover the increase in activity in
the first few Myr observed at a given mass (Sect. \ref{sect:conc_XvsA}) would
imply, in this  hypothesis, a spin-down of stars in the early PMS. Although
theoretical speculations suggest that this could indeed happen due to
non-instantaneous disk locking \citep{har02}, observational studies of the
rotational evolution in the ONC and NGC~2264 \citep{reb02a} rather suggest that
stars keep constant period as they evolve down the convective tracks. Given the
lack of firm observational evidence of the super-saturation regime, here we do
not further explore this hypothesis but rather present and discuss an
alternative explanation.

\subsection{The disk hypothesis}

We now introduce a simple model that tries to account for the observed
evolution of activity of low mass stars in the early PMS phase. Given the
persistent lack of rotational data for the greatest part of our stellar sample,
the uncertainties in the X-ray data and in the theoretical evolutionary models,
we keep our model extremely simple and, at this point, consider it only as a
hint for further observational and theoretical work.

The observational and theoretical bases of our model are the followings:

\begin{itemize}

\item Most PMS stars, especially in the older regions and/or not of very low
mass have almost saturated activity levels: $\log(L_{\rm X}/L_{\rm bol}) \sim -3.0$. A
relation between activity and rotation is generally not observed. As discussed
at the end of the last section, the relation of $L_{\rm X}/L_{\rm bol}$ with Rossby number
(Eq. \ref{eq:concR0}) suggests that activity should indeed be saturated.

\item Stars with circumstellar accretion disks, in several star forming regions,
show lower median activity levels (both in $L_{\rm X}$ and $L_{\rm X}/L_{\rm bol}$) respect to
those with weaker or no evidence of a disk and/or accretion
\citep{stel01,fla02b,fla02c}. The origin of the lower activity level is at
present unknown.

\end{itemize}

We will make the following assumptions:

1) Departure from saturation is due exclusively to the presence of disks, stars
that have already shed their disks having saturated $\log(L_{\rm X}/L_{\rm bol})$.

\begin{equation}
H_{\rm sat.} \equiv \log(L_{\rm X}/L_{\rm bol})_{\rm sat.}=-3
\end{equation}

2) Stars with a disk all have the same value of $\log(L_{\rm X}/L_{\rm bol})$ which we
will  take to  be -5.0.

\begin{equation}
H_{\rm disk} \equiv \log(L_{\rm X}/L_{\rm bol})_{\rm disk}=-5
\end{equation}

Although this value is quite arbitrary, it is compatible with the lowest
values of $\log(L_{\rm X}/L_{\rm bol})$ of low mass ($< 3M_\odot$) ONC and $\rho$ Ophiuchi
stars. According to maximum likelihood distributions, only  $\sim$20\% of both
the high accretion ONC stars (as defined by \citealt{fla02b}) and of all $\rho$
Ophiuchi members in our sample, have $\log(L_{\rm X}/L_{\rm bol})$ below this threshold.
Note also that in a more realistic model which allows for a continuous range of
$L_{\rm X}/L_{\rm bol}$ values and of disk evolutionary stages, $H_{\rm disk}$ would
correspond to $\log(L_{\rm X}/L_{\rm bol})$ of stars with the least evolved accretion
disks.

3) The dissipation of circumstellar disks can be described by a stochastic
process with characteristic time-scale $\tau_{\rm disk}$; we can express the
fraction of stars that retain their disk, $f_{\rm disk}$, at a time $t$ by:

\begin{equation}
f_{\rm disk}=\frac{N_{\rm disk}}{N_{tot}}=e^{-t/\tau_{\rm disk}}
\end{equation}

We can then compute the mean $\log(L_{\rm X}/L_{\rm bol})$ as a function of $t$, $<H(t)>$.


\begin{equation}
<H(t)> = H_{\rm sat.} + (H_{\rm disk} - H_{\rm sat.}) \ e^{-t/\tau_{\rm disk}}
\end{equation}

This model has two unknown parameters: $H_{\rm disk}$, which we have fixed at
-5, and $\tau_{\rm disk}$ which we keep as a free parameter. $H_{\rm sat}$ is
here considered as fixed  by the observations at -3, roughly corresponding to 
the highest values observed in the PMS and equal to the saturation level
obtained for fast rotating MS stars. The right column of Fig. 
\ref{fig:lx_lxlb_age} shows, superposed on the already described observed
trends of $\log(L_{\rm X}/L_{\rm bol})$ with age for different mass ranges, three
realization of the model corresponding to three values of the disk dissipation
time-scale, 1.0, 1.5 and 2.0 Myrs. We note two facts: 

1) The agreement between data for the PMS regions and models is qualitatively
good, also considering the extreme simplicity of our assumptions. The models 
do not, and are not supposed to, reproduce the fall in $L_{\rm X}/L_{\rm bol}$  on the
MS, which can be attributed either to wind-driven spin down of the stars or,
for the $2-3M_\odot$ bin, to the loss of the convective region.

2) The observed evolution of $L_{\rm X}/L_{\rm bol}$ is somewhat faster for more massive
stars. The model seems to suggest that $\tau_{\rm disk}$ depends on stellar
mass, ranging from $\sim 2.0$ Myrs for $M \sim 0.2 M_{\rm \odot}$ to $\sim 1.0$
Myrs for $M \sim 0.75 M_{\rm \odot}$, where the quoted masses are simply the
centers of the second and fourth mass bins.

We therefore seem to have derived an indication on the circumstellar disk
lifetime. How do these results compare with totally independent results
obtained from specific studies performed at infrared wavelengths?
\citet{hai01a} analyze a sample of six clusters and determine, from a linear
fit of disk fraction vs. cluster age, a disk lifetime of about 6Myr, this being
the age at which essentially all the stars loose their near-IR detectable
disks. A similar analysis performed by \citet{hil03} using a larger sample of
star forming regions, suggests instead an inner disk lifetime of only 2-3 Myr.
An independent indication on the timescale for the evolution of inner disks has
been recently  derived by \citet{her03} from the rotational periods of ONC
members. Interpreting the period distribution in the context of the disk locking
hypothesis, they conclude that the timescale for ONC stars to be released from
their disk locking is $0.5-1.0$ times the ONC age. If the two median stellar
ages reported in Fig.  \ref{fig:conc_SFR_age} for our ONC sub-regions are used
this translates to $0.7-2.0$ Myr.


Our best guess for $\tau_{\rm disk}$ derived from X-ray data, i.e. 1-2Myrs.,
falls within the range of the above estimates. We note that all these estimates
are  on one hand subject to significant uncertainties, and on the other hand may
trace different, though related phenomena such as the presence of the inner
disk, disk locking and mass accretion. The order of magnitude agreement is
therefore suggestive. Although we have identified the evolutionary time-scale
of $L_{\rm X}/L_{\rm bol}$ with the disk dissipation time-scale, our model is actually
rather general: it only assumes a two level population and a fixed transition
probability per unit time from the low to the high level. We have tentatively
identified the low level with the presence of disks because recent studies
\citep{fla02b,fla02c} indicate that stars surrounded by accretion disks have
lower levels of $\log(L_{\rm X}/L_{\rm bol})$. If the lowered activity level is in fact
due to accretion itself as opposed to the presence of a disk, the time-scale
for the evolution of activity might correspond to the time-scale for the
evolution of accretion, which might be shorter than the time-scale for the
disappearance of IR detectable disks. The rather good agreement of our X-ray
derived time scale with the disk locking times scale reported by \citet{her03}
is particularly suggestive as disk locking implies a modification of the
magnetic field geometry, conceivably affecting the X-ray emitting plasma. 

As for our second result, i.e. the dependence of the X-ray derived $\tau_{\rm
disk}$ from stellar mass, we note that this dependence is also suggested, and
in the same direction found here, by several IR studies (e.g.
\citealt{hai01b,hil98}), thus giving further support to our interpretation. 

More X-ray and optical/IR data are needed to fully explore the relation between
magnetic activity and accretion and/or disks.

\section{Summary and Conclusions \label{sect:conc_sum}}

We have studied the evolution of  X-ray activity in the first stages of stellar
evolution, from the early PMS ($\sim 1$ Myr.) to the ZAMS ($\sim 100$ Myr),
using published and archival data on five star forming regions and two young
ZAMS clusters. In all cases we have reviewed the X-ray and optical data used to
characterize members of the respective regions. If necessary, we have repeated
the data analysis in order to homogenize and update, to the best of our
knowledge, the assumptions used to derive activity indicators ($L_{\rm X}$ and
$L_{\rm X}/L_{\rm bol}$) and stellar parameters (masses and ages). 

In the early PMS phase we observe that low mass stars that posses a convective
region show high levels of activity, often close to the saturation level
($\log \ L_{\rm X}/L_{\rm bol} \sim -3$) found for fast rotating MS stars. We compare these
activity levels with those predicted by two physically distinct scenarios:

{\bf \#1}) the total energy output of the corona ($L_{\rm X}$) is determined
exclusively by stellar rotation ($P_{\rm rot}$) and saturates when $L_{\rm X}$ reaches
$10^{-3}$ times the total stellar luminosity.

{\bf\#2}) the fraction of energy output by the corona respect to the whole
energy produced by the star is determined by the dimensionless ratio between
the characteristic time of rotation and the characteristic time of convection
(i.e. the Rossby number, $R_{\rm 0}$). This relation saturates at a fixed value of
$R_{\rm 0}$, corresponding to $L_{\rm X}/L_{\rm bol}\sim 10^{-3}$.

While, for MS stars, these two scenarios are hardly distinguishable, both
giving  a fair description of the observations, they predict distinctively
different results if applied to $1-5$ Myr old stars. Scenario \#1, in order to
explain the high levels of activity observed, would require very fast
rotations, which are simply not observed. Scenario \#2 on the other hand
predicts that stars in the early PMS with $P_{\rm rot} \lesssim 16$ d, i.e.
consistent with the rotations measured in the ONC, should all be saturated.

Our data seem therefore to favor scenario \#2, although it does not fully
account for the observations: median $L_{\rm X}/L_{\rm bol}$ appears to be somewhat lower
than the saturation value for young ages ($\lesssim 5$Myr) and low masses
($\lesssim 1M_\odot$), but tends to reach that level at older ages and larger
masses. Saturation is then retained up to the age of the Pleiades and NGC~2516
($100-140$Myrs) for stars with $M \lesssim 0.5M_\odot$, while activity in more
massive stars seems to desaturate earlier. We can tentatively explain this
latter decay of activity from the PMS to the MS with the decrease of the
convective turnover time (or, equivalently, convective zone depth), which is
indeed more pronounced for more massive stars \citep{ven98} and, possibly, with
the angular momentum loss induced by magnetic winds (cf. \citealt{bou97}). The
increase of $L_{\rm X}/L_{\rm bol}$ in the early PMS however is not so readily explained
by the foreseen evolution of the Rossby number. Supported by observational
evidence, we put forth a simple model that assumes the existence of two
populations: one in which activity is at the saturation level, the other with
lower observed activity, which we tentatively identify with stars surrounded by
circumstellar disks and/or undergoing accretion. We then postulate that the low
activity population makes a transition to the high activity population (e.g.
dissipates disks or stops accreting) with a characteristics time, $\tau$.
Observed median activity levels are roughly consistent with such a model if
$\tau \sim 1-2$ Myr, being at the short end of the interval for higher mass
stars ($\sim 1 M_\odot$)  and at the long end for the lowest mass ones ($\sim
0.2 M_\odot$). These characteristic times are similar to disk dissipation
times derived by IR studies, to the timescale for disk locking, as derived by
modeling the rotational periods distribution of ONC stars, and to the expected
timescale for the evolution of circumstellar accretion. Given the considerable
uncertainties in all of these estimates, we restrain from any more precise
identification of the physical process that determines the evolution of
activity. We also note that the observed trend of $\tau$ with mass is
consistent, according to recent IR studies, with the disk (and possibly with
the accretion and disk locking) hypothesis.

\begin{acknowledgements} The authors are thankful to P. Ventura for providing
his model calculation of convective turnover times in the PMS and to the
referee, T. Preibisch, for comments that helped improve this work. E.F. wish to
thank F. Damiani for providing the X-ray data for NGC~2516 prior to publication
and N. Pizzolato for useful discussions. The authors wish to acknowledge
support from the Italian Space Agency (ASI) and MURST.  \end{acknowledgements}

\end{document}